\documentclass[aps,prd,nofootinbib,twocolumn,superscriptaddress,floatfix,notitlepage]{revtex4-2}

\usepackage{graphicx}
\usepackage{amsmath,amsfonts,amssymb}
\usepackage{color}
\usepackage{verbatim}
\usepackage{enumitem}
\usepackage{aas_macros}
\usepackage[breaklinks,colorlinks,urlcolor=blue,citecolor=blue,linkcolor=magenta]{hyperref}

\newcommand{\td}{{\rm d}}
\newcommand{\vect}[1]{\boldsymbol{#1}}

\newcommand{\be}{\begin{equation}}
\newcommand{\ee}{\end{equation}}
\newcommand{\bea}{\begin{equation} \begin{aligned}}
\newcommand{\eea}{\end{aligned} \end{equation}}

\def\lsim{\mathrel{\raise.3ex\hbox{$<$\kern-.75em\lower1ex\hbox{$\sim$}}}}
\def\gsim{\mathrel{\raise.3ex\hbox{$>$\kern-.75em\lower1ex\hbox{$\sim$}}}}

\newcommand{\papertitle}{Constraints on Dark Matter Models from Supermassive Black Hole Evolution}

\begin{document}

\title{\papertitle}

\author{John Ellis}
\email{john.ellis@cern.ch}
\affiliation{King’s College London, Strand, London, WC2R 2LS, United Kingdom}
\affiliation{Theoretical Physics Department, CERN, Geneva, Switzerland}
  
\author{Malcolm Fairbairn}
\email{malcolm.fairbairn@kcl.ac.uk}
\affiliation{King’s College London, Strand, London, WC2R 2LS, United Kingdom}
   
\author{Juan Urrutia}
\email{juan.urrutia@kbfi.ee}
\affiliation{Keemilise ja Bioloogilise F\"u\"usika Instituut, R\"avala pst. 10, 10143 Tallinn, Estonia}
\affiliation{Department of Cybernetics, Tallinn University of Technology, Akadeemia tee 21, 12618 Tallinn, Estonia}

\author{Ville Vaskonen}
\email{ville.vaskonen@pd.infn.it}
\affiliation{Keemilise ja Bioloogilise F\"u\"usika Instituut, R\"avala pst. 10, 10143 Tallinn, Estonia}
\affiliation{Dipartimento di Fisica e Astronomia, Universit\`a degli Studi di Padova, Via Marzolo 8, 35131 Padova, Italy}
\affiliation{Istituto Nazionale di Fisica Nucleare, Sezione di Padova, Via Marzolo 8, 35131 Padova, Italy}

\begin{abstract}
A semi-analytical model for the evolution of galaxies and supermassive black holes (SMBHs) within the $\Lambda$CDM paradigm has been shown to yield stellar mass-BH mass relations that reproduce both the JWST and pre-JWST observations. Either fuzzy or warm dark matter (FDM or WDM) would suppress the formation of the smaller galactic halos that play important roles in the CDM fit to the high-redshift SMBH data. Our analysis of the stellar mass-BH mass relation disfavours FDM fields with masses $< 2.0\times 10^{-20}$~eV and WDM particles with masses $< 7.2$~keV, both at the 95\% confidence level.
\\~~\\
KCL-PH-TH/2025-37, CERN-TH-2025-186, AION-REPORT/2025-07
\end{abstract}

\maketitle

\noindent\textbf{Introduction  -- } The advent of the James Webb Space Telescope (JWST) has enabled surveys to probe the infrared universe with unprecedented accuracy~\cite{2025arXiv250104085F,2024MNRAS.533.3222D,2024ApJ...965L...6B,2023ApJ...954...31C,2023arXiv230602465E,2023ApJS..269...16R,Ostlin:2024rkp,2024ApJ...974...92B}, facilitating exploration of the reionization era and before (see~\cite{2024arXiv240521054A} for a review). One of the most interesting developments has been new insight into proposed mechanisms for the formation of supermassive black holes (SMBHs)~\cite{Woods:2018lty,Inayoshi:2019fun,Volonteri:2021sfo}. It had been suggested that they could have been seeded by black holes (BHs) with masses ${\cal O}(10^2 - 10^3) M_\odot$ that could be the remnants of population~III (Pop-III)~\cite{2025MNRAS.536..851C} stars that collapsed at redshifts $z\gsim 10$ (light-seed scenario), or by heavier BHs with masses ${\cal O}(10^5) M_\odot$ in the cores of protogalaxies that subsequently merged to form the larger galaxies seen at lower redshifts (heavy-seed scenario)~\cite{Bromm:2002hb,Lodato:2006hw}.

Using JWST, a new population of high-$z$ BHs has been discovered~\cite{2023Natur.621...51D,2023A&A...677A.145U,2023ApJ...953L..29L,2023ApJ...959...39H,Bogdan:2023ilu,Maiolino:2023bpi,2024ApJ...966..176Y,2023ApJ...954L...4K,2024ApJ...964...90S} that represent a challenge for the conventional picture of SMBH formation~\cite{Pacucci:2023oci,Ellis:2024wdh,2025ApJ...981...19L,2025arXiv250608116D}. The high-$z$ SMBHs seem to require either a heavy-seed scenario~\cite{Natarajan:2023rxq,Toubiana:2024bil,2024arXiv240403576K} or a very high efficiency for mergers of light seeds~\cite{Ellis:2024nzv,2024MNRAS.531.4311B} in order for them to be a viable option. The possibility of growth through mergers inside nuclear clusters has also been tested against the JWST results~\cite{Kritos:2024sgd}. An active topic of debate has been whether these SMBHs are accreting at super-Eddington rates, in which case their masses may have been overestimated~\cite{2024MNRAS.530.1732P,Lupi:2024cxv,Lambrides:2024ugh,Madau:2025kpn,Zana:2025uuk}. In parallel, the discovery of `little red dots'~\cite{2023ApJ...951L...5Y,2024arXiv240906772T,2024ApJ...964...39G,Matthee:2023utn} has defied many astrophysical assumptions about galaxies at high $z$~\cite{Shen:2024hpx,Gentile:2024uiw,Kokubo:2024ukw}. There is an active debate centered around the nature of these objects and the role that SMBHs play in them~\cite{Ananna:2024jug,Madau:2024fdv,Pacucci:2024tws,Naidu:2025rpo,LRDmass}. 
 
This work explores the evolution of SMBH seeds in CDM, FDM, and WDM universes, building on our earlier results for the star formation rate~\cite{Ellis:2025xju} and our semianalytical model for the coevolution of galaxies and SMBHs~\cite{Ellis:2024nzv}. We compare model predictions to the stellar mass-BH mass observations across a broad redshift range and conclude that, despite its uncertainties, SMBH evolution is a sensitive probe of DM properties, indeed more powerful than the stellar UV luminosity functions that we studies previously~\cite{Ellis:2025xju}. The reason is that the bright stars that dominate the UV emission are short-lived, so the UV luminosity function serves as a cosmic ``snapshot'' in which the lightest halos are too dim to be probed. In contrast, SMBHs are long-lived, carrying information from as early as $z\sim20$ when they were first formed in light halos, which are very sensitive to the nature of DM. 

We show here that the best-fit parameters for the evolution of the SMBH population in the CDM model are well within the ranges expected from simulations and physical arguments, making the observed SMBH population less surprising than previously thought. We further conclude that the current population of SMBHs disfavours FDM fields with masses $< 2.0\times 10^{-20}$~eV and WDM particles with masses $< 7.2$~keV, both at the 95\% confidence level. Finally, for the light-seed scenario to be viable, it requires the buildup of light halos, making the light-seed case more sensitive to the nature of DM. This correlation between the nature of DM and the origin of SMBHs, will provide an interesting phenomenology for future gravitational wave detectors like LISA~\cite{LISA:2024hlh} or AION~\cite{AION:2025igp}, which we plan to explore in future work.

\vspace{5pt}\noindent\textbf{Methodology -- } For each DM model, we begin by embedding a seed of mass $m_{\rm seed}$ at some redshift $z_{\rm seed}$ in every halo that is heavier than some minimal mass $M_{\rm seed}$, and evolve the BH masses and stellar masses numerically using the formalism derived in Ref.~\cite{Ellis:2024nzv,Dayal:2018gwg} and summarized in the Supplemental Material~\cite{Ellis2026}. Our model includes the 8  parameters described in Table~\ref{tab:param}. We resolve halos in the range $M_{\rm v}=10^2-10^{16}\,M_{\odot}$ and plant the seeds at $z_{\rm seed} = 20$. We have checked that our results are insensitive to this choice, and that the redshift steps and halo mass bins are small enough that the solutions are not affected.\footnote{The complete SMBH-galaxy coevolution from $z_{\rm seed}$ to $z=0$ takes a few seconds in Mathematica on a M2 MacBook Air.} 

\begin{table}
    \centering
    \begin{tabular}{c|c}
        Parameter & Interpretation \\
        \hline
        $m_{\rm seed}$ & Mass of the SMBH seed.\\
        \hline
        $M_{\rm seed}$ & Lowest halo mass that is planted with seeds.\\
        \hline
        $p_{\rm BH}$ & The probability that a halo merger ends \\
        &in the merger of the central SMBHs. \\
        \hline
        $f_{\rm SN}$& The fraction of gas \\
        &that is not expelled after SN feedback.\\
        \hline
        $f_{\rm Edd}$ & The maximal allowed accretion \\
        & as a fraction of the Eddington limit.\\
        \hline
        $f_1^{\rm acc}$& The fraction of incoming baryons \\
        &accreted onto the SMBH.\\
        \hline
        $f_2^{\rm acc}$& The rate of accretion of baryons in the halo.\\
        \hline
        $m_{\rm FDM/WDM}$ & DM mass, if not CDM.\\
    \end{tabular}
    \caption{List of the parameters in our model and their interpretations.}
    \label{tab:param}
    \vspace{-0.5 cm}
\end{table}

\begin{figure}
    \centering
    \includegraphics[width=0.9\columnwidth]{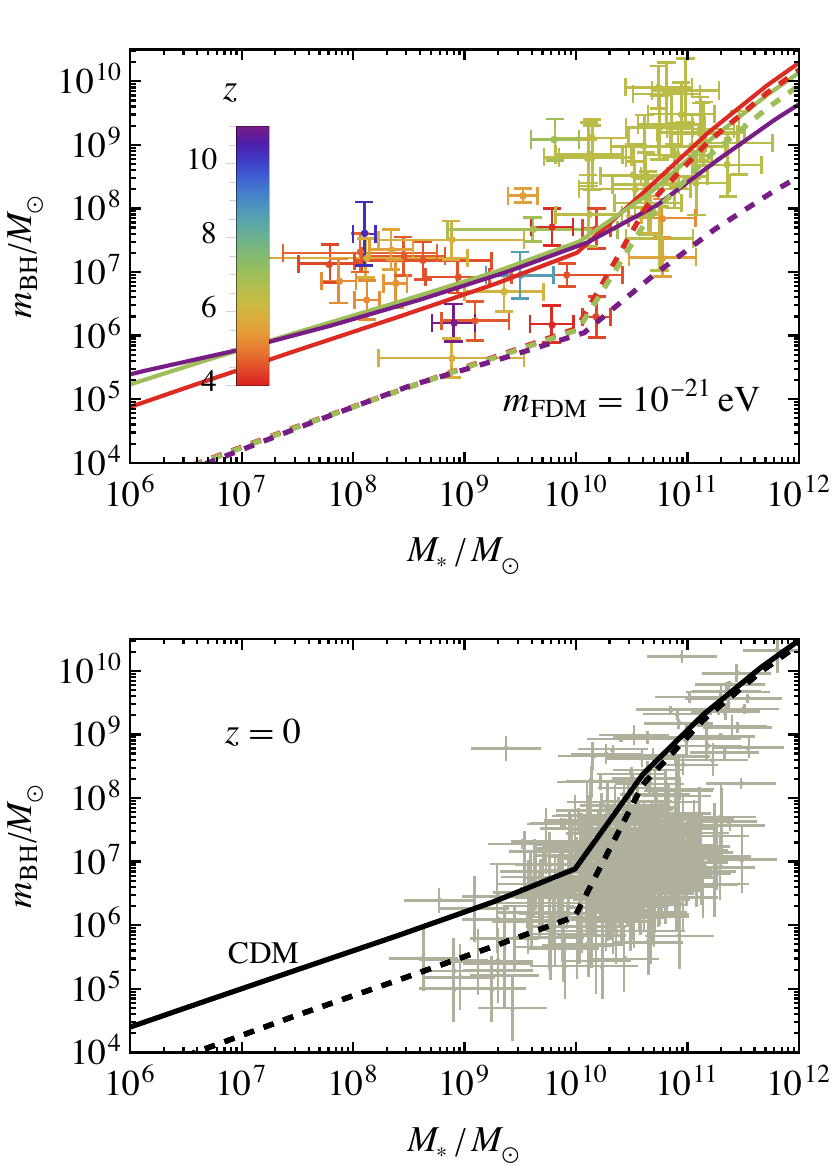}
    \vspace{-0.2 cm}
    \caption{Stellar mass-BH mass relation in a heavy-seed scenario. Solid lines show the CDM results and dashed lines the FDM result with $m_{\rm FDM}=10^{-21}\,{\rm eV}$. The different colours in the top panel represent the different redshifts, while the lower panel represents the local universe.}
    \label{fig:scalings}
\end{figure}

In Fig.~\ref{fig:scalings} we show the resulting stellar mass-BH mass relation at different redshifts for a benchmark case with $\log_{10}(m_{\rm seed}/M_{\odot}) = 5$, $\log_{10}(M_{\rm seed}/M_{\odot}) = 7$, $\log_{10}f_{\rm SN} = -3.0$, $\log_{10}f_1^{\rm acc} = -19.5$, $\log_{10}(f_2^{\rm acc}{\rm Myr}) = -3.0$, $\log_{10}f_{\rm Edd} = 0.5$ and $p_{\rm BH} = 0.1$. We display results for two different cosmologies, CDM and FDM, with $m_{\rm FDM} = 10^{-21}\,{\rm eV}$. We see in the lower panel that both DM models can approximately reproduce the observations at $z=0$, but at high $z$, shown in the upper panel, the prediction is very sensitive to the DM model. There are two reasons for this. First, the abundance of low-mass halos is suppressed at the time of seeding for the non-CDM models,\footnote{{Our implementations of the halo mass function cutoffs for FDM and WDM are described in detail in Section 3.1 of~\cite{Ellis:2025xju}. The implementation for FDM is based on~\cite{Hu:2000ke} and~\cite{Passaglia:2022bcr}, whose fitting formula we use. The implementation for WDM is based on~\cite{Bode:2000gq,Hansen:2001zv,Viel:2005qj}. The DM properties can also affect the formation of seeds. In particular, direct collapse SMBH seeds can form in WDM models at lower redfhifts than in the CDM model~\cite{Dayal:2017yhb}. We do not account for this effect but use a generic parametrization of the seeds.}} and, second, at these masses, the SN feedback during reionisation makes accretion inefficient, so mergers at very high redshift can be the main channel for the growth of these SMBHs~\cite{2025arXiv250609184B}, which is further suppressed. The effect of WDM (not shown) is qualitatively similar to that of FDM. This example illustrates how the suppression of small-scale structure abundance affects the evolution of SMBHs.

\begin{figure}
    \centering
    \includegraphics[width=0.9\columnwidth]{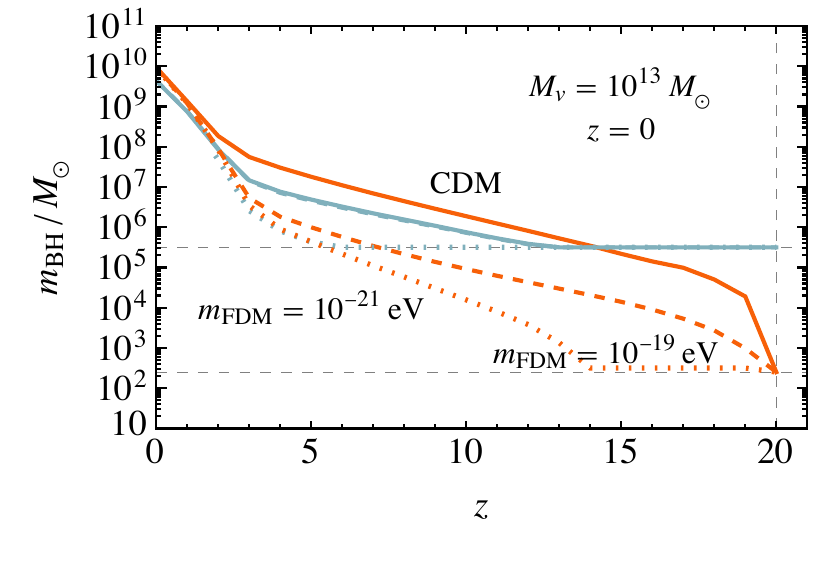}
    \vspace{-0.5 cm}
    \caption{Growth of individual SMBHs in both the light and heavy seed scenarios (orange and blue, respectively). CDM calculations are shown with solid curves and FDM with dashed ($m_{\rm FDM} = 10^{-19}$\,eV) and dotted ($m_{\rm FDM} = 10^{-21}$\,eV) curves. The host DM halo mass at $z=0$ is fixed to $M_{\rm v} = 10^{13}\,M_\odot$.}
    \label{fig:single_growth}
\end{figure}

In Fig.~\ref{fig:single_growth} we show the average evolution of an SMBH in a DM halo whose mass today is $M = 10^{13} M_\odot$. The blue curves show a heavy seed case, with the same values as for Fig.~\ref{fig:scalings}, and the orange curves are for a light seed case where we have taken $\log_{10}(m_{\rm seed}/M_{\odot}) = 2.5$, $\log_{10}(M_{\rm seed}/M_{\odot}) = 4.5$ and $p_{\rm BH}=1$. The solid, dashed and dotted curves correspond, respectively, to CDM, FDM with $m_{\rm FDM}=10^{-19}\,{\rm eV}$ and FDM with $m_{\rm FDM}=10^{-21}\,{\rm eV}$. We see that the growth curves are similar at low redshifts and that the heavy seed case grows steadily from $z = 12$, whereas in the CDM model, the light seed case shows more rapid growth at the beginning. The reason is that a large number of light halos are simultaneously trapped in a collapsing overdensity that becomes a heavy halo.\footnote{We treat the mergers as instantaneous. In a more realistic model, this process will take time. We plan to explore this further in future work.} When looking at the dashed and dotted curves, we see that the FDM model leads to a slower SMBH growth at high-$z$ than the CDM model, but that the predictions for low-$z$ are very similar. The light-seed case is more sensitive to the DM model since it relies more on the merging of light halos for the build-up of the seeds. If lighter FDM masses are considered, the same problem affects the heavy-seed case, as shown by the dotted curve.

\vspace{5pt}\noindent\textbf{Results  -- } We perform a Markov-chain Monte Carlo (MCMC) analysis over the model parameters $\vect{\theta}$ (see Table~\ref{tab:param}), comparing the predictions in the $m_{\rm BH}\, ,M_{*}$ plane at each redshift to the observations~\cite{2023Natur.621...51D,2023A&A...677A.145U,2023ApJ...953L..29L,2023ApJ...959...39H,Bogdan:2023ilu,Maiolino:2023bpi,2024ApJ...966..176Y,2023ApJ...954L...4K,2024ApJ...964...90S,2015ApJ...813...82R,2021ApJ...914...36I}. In each case, we generate four MCMC chains of $2\times10^4$ steps with $5\times10^3$ steps burn-in, and check the convergence of the chains using the Gelman-Rubin statistic~\citep{Gelman:1992zz}. The likelihood is given by
\bea
    &\mathcal{L}(\vect{\theta}) \propto \prod_j \! \int \!\td \log_{10} \!m_{\rm BH}  \td \log_{10} M_*  \frac{\td P(m_{\rm BH}|M_*,z_j,\vect{\theta})}{\td \log_{10} m_{\rm BH}}\\ 
    &\qquad\qquad\times\mathcal{N}(\log_{10} m_{\rm BH} | \log_{10} m_{{\rm BH},j}, \sigma_{{\rm BH},j})\\
    &\qquad\qquad\times\mathcal{N}(\log_{10} M_*|\log_{10} M_{*,j},\sigma_{*,j}) \,,
\eea
where $j$ labels the data points, the Gaussian probability densities $\mathcal{N}(x|\bar{x},\sigma)$ account for the uncertainties in the measurements of the BH masses and the galaxy stellar masses and $\td P(m_{\rm BH}|M_*,z_j)/\td \log_{10} m_{\rm BH}$ denotes the occupation fraction of BH masses in galaxies with stellar mass $M_*$ at redshift $z_j$, which we obtain from the numerical computation described in the previous section. For each parameter, we consider log-uniform priors. 

When analysing the stellar mass-BH mass relation in CDM we see a strong correlation between $m_{\rm seed}$ and $M_{\rm seed}$, which is qualitatively compatible with the findings of our previous work~\cite{Ellis:2024nzv}. The full posteriors can be seen in the Supplemental Material~\cite{Ellis2026} in Fig.~\ref{fig:scan_CDM}. Our results for the CDM case are similar also to those of~\cite{Jeon:2025rqt}: A wide range of model parameters can reproduce the low-$z$ SMBH population, while the discrimination between different scenarios becomes significantly more pronounced at redshifts $z \gtrsim 10$. Both the light- and heavy-seed scenarios can provide good fits, but for light seeds of $m_{\rm seed} < 10^4\,M_{\odot}$ they need to be planted in very light halos, $M_{\rm seed} \lsim 10^6\,M_{\odot}$. Whether such initial conditions are physical is up for debate.\footnote{The results in this paper are rather insensitive to the star formation rate at high $z$ and, in particular, would not depend strongly on any change in Pop-III production.} Some studies have found such light halos to be able to host Pop-III stars~\cite{2007MNRAS.377..667N} that could provide the BH seeds, although the majority of studies find higher halo mass thresholds~\cite{2021MNRAS.507.1775S,2021ApJ...917...40K}. The range of predictions from these studies falls within our $68.5\,\%$ confidence level (CL) region.

We found that light seeds require high merger efficiency $p_{\rm BH}\simeq 1$, whereas for heavy seeds, the preference is $p_{\rm BH}\simeq 0.1$. Low values of $p_{\rm BH} \lsim 0.1$ are excluded at the 68.5 \%  CL. This range, $p_{\rm BH} \gtrsim 0.1$, is consistent with that required to reproduce the observed PTA gravitational-wave background~\cite{Ellis:2023dgf}. If SMBH binaries are driven solely by gravitational wave emission, the PTA data prefer $p_{\rm BH} \approx 0.17$~\cite{Ellis:2023dgf}, which, as mentioned above, would indicate a preference for heavy seeds.

We find an upper bound of $f_{\rm SN}<10^{-2.21}$ at the $68.5\%$  CL on the efficiency of the SN feedback. If $f_{\rm SN}$ is too large, the SMBHs can accrete efficiently at all masses. This results in a power-law stellar mass-SMBH mass relation that does not give a good fit for the local AGNs, as has been seen in simulations that suppress the SN feedback~\cite{2017MNRAS.465...32B}. The other accretion parameters $f_1^{\rm acc}$, $f_2^{\rm acc}$ and $f_{\rm Edd}$ are less constrained. We find that, as expected, there is a preference for either Eddington or super-Eddington growth. The correlation with the seed mass is not as strong as with $p_{\rm BH}$, but we find that the heavy seed case does not have a preference for super-Eddington growth. 

\begin{figure*}
    \centering
    \includegraphics[width=0.9\textwidth]{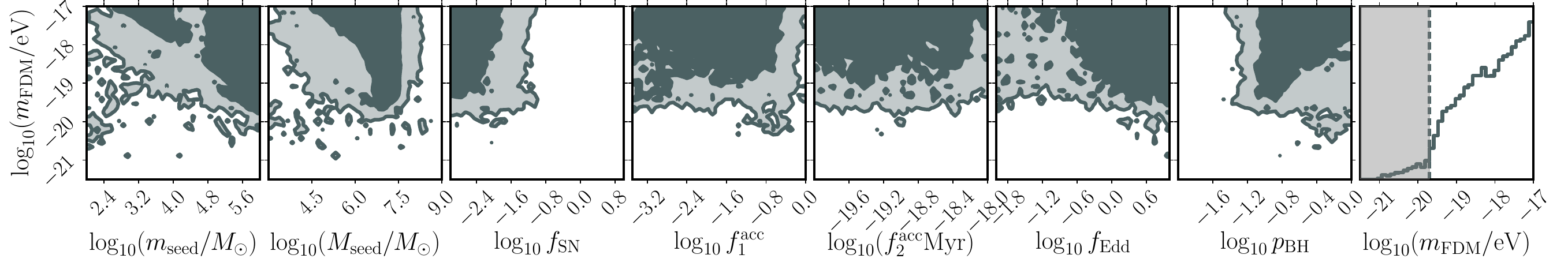}
    \includegraphics[width=0.9\textwidth]{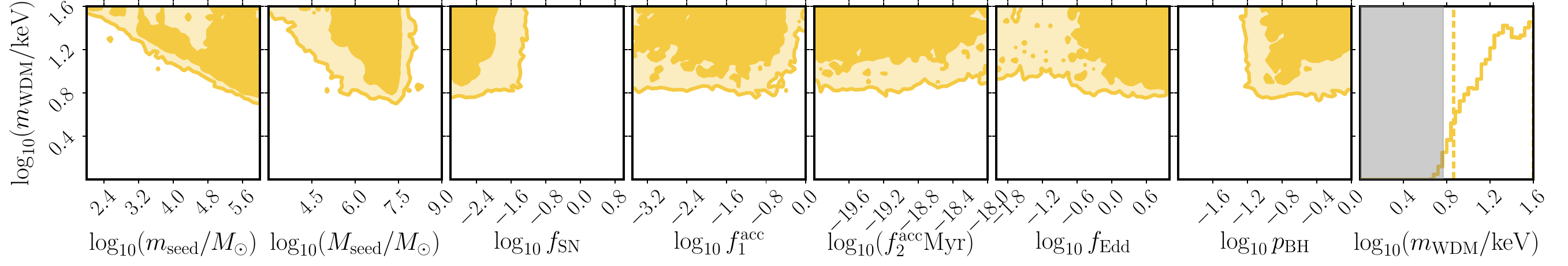}
    \caption{Posteriors of the stellar mass-BH mass fit parameters as functions of the FDM mass (upper panels) and WDM mass (lower panels). The vertical dashed lines are our 95\% CL lower bouds on these masses. The grey band in the $\log_{10}(m_{\rm FDM}/{\rm eV})$ posterior shows the most stringent 95$\%\, {\rm CL}$ exclusion of $m_{\rm FDM}$ from a Lyman-$\alpha$ analysis~\cite{Irsic:2017yje}, namely $< 2\times10^{-20}\,{\rm eV}$, and the grey band in the $\log_{10}(m_{\rm WDM}/{\rm keV})$ posterior shows the most stringent 95$\%\, {\rm CL}$ exclusion of $m_{\rm WDM}$ from a joint analysis of strong gravitational lensing, the Lyman-$\alpha$ forest and Milky Way satellites~\cite{Enzi:2020ieg}, namely $< 6.0\,{\rm keV}$. The full corner plots are shown in the Supplemental Material~\cite{Ellis2026}.}
    \label{fig:DM_post}
\end{figure*}

We perform similar analyses of the FDM and WDM cases, in which we consider the effect on structure evolution. We do not consider changes in the accretion model~\cite{Chiu:2025vng}. The resulting posteriors are shown in Fig.~\ref{fig:DM_post}. We find lower bounds on the FDM and WDM masses that depend on the SMBH seed mass: The lighter the seeds are, the higher the FDM or WDM mass has to be. The same correlation between $m_{\rm seed}$ and $p_{\rm BH}$ that we found in the CDM case holds here, and the smaller the merging efficiency, the less stringent the DM mass bounds are. Marginalizing all parameters, we find that $m_{\rm FDM} < 2.0\times10^{-20}\,{\rm eV}$ and $m_{\rm WDM} < 7.2\,{\rm keV}$ are excluded at the 95\% CL. If the correct seeding scenario is light seeds, something that could be tested with future gravitational wave detectors~\cite{AION:2025igp}, then the bounds would become much more stringent.

In the case of FDM, the bound is as stringent as the bound coming from Lyman-$\alpha$ clouds, which implies $m_{\rm FDM} > 2 \times 10^{-20}$~eV ~\cite{Rogers:2020ltq}. Other cosmological and astrophysical bounds include $m_{\rm FDM} > 0.6 \times 10^{-19}$~eV from Eridanus II~\cite{Marsh:2018zyw}, $m_{\rm FDM} \gsim 3 \times 10^{-19}$~eV from Segue 1 and 2~\cite{Dalal:2022rmp}, $m_{\rm FDM} \gsim 2.2 \times 10^{21}$~eV from Cetus-II~\cite{Benito:2025xuh}, and $m_{\rm FDM} > 4.4 \times 10^{-21}$~eV from gravitational lensing~\cite{Powell:2023jns}. Some works suggest evidence for FDM with $m_{\rm FDM} \lsim 10^{-22}$~eV~\cite{Amruth:2023xqj,Rogers:2023upm}, which conflicts with these limits. As a function of the SMBH seed mass, the lower bounds are approximately $m_{\rm FDM} \gsim 10^{-20}\,{\rm eV} \,(m_{\rm seed}/10^6 M_\odot)^{-3/4}$. 

As seen in Fig.~\ref{fig:DM_post}, we find a very similar behaviour in the WDM case. The current SMBH population implies $m_{\rm WDM}>7.2\,{\rm keV}$, which is slightly more stringent than the bound coming from a combined analysis of strong gravitational lensing, Lyman-$\alpha$ clouds and Milky Way satellites~\cite{Enzi:2020ieg} of $m_{\rm WDM} > 6.0$~keV. Again we see a correlation with the seeding mass of the form {$m_{\rm WDM} \gsim 10^{0.7}\,{\rm keV} \,(m_{\rm seed}/10^6 M_\odot)^{-0.225}$}. Other constraints include: $m_{\rm WDM} > 5.3$~keV from Ly$\alpha$ data~\cite{Irsic:2017ixq}, $m_{\rm WDM} > 0.8$~keV from lensing at $z \sim 0.2$~\cite{Vegetti:2018dly}, $m_{\rm WDM} > 5.6$~keV from quasar lensing~\cite{Hsueh:2019ynk}, $m_{\rm WDM} > 3.0$~keV from quadruple-image lenses~\cite{Gilman:2019nap}, $m_{\rm WDM} > 3.6$~keV from Milky Way satellites~\cite{Dekker:2021scf}, and $m_{\rm WDM} > 3.2$~keV from JWST UV luminosity data~\cite{Liu:2024edl}.

\vspace{5pt}\noindent\textbf{Conclusions  -- } We have found that SMBHs could be better discriminators of DM scenarios than stars~\cite{Ellis:2025xju}. The reason is that, while stars are short-lived, SMBHs live practically forever. Consequently, the SMBH population retains imprints of their seeding scenario. Since seeding occurred at very high redshifts in low-mass halos, these imprints are ideal for testing deviations from CDM.

Leveraging this insight, we have used the latest data from the JWST on both the stellar population, which informs the internal galactic processes, and the masses of the galaxies that host the SMBH, in conjunction with previous data, to put new bounds on DM. We have found that FDM masses of $m_{\rm FDM} < 2.0\times10^{-20}\,{\rm eV}$ and WDM masses of $m_{\rm WDM}<7.2\,{\rm keV}$ are excluded at the $95\%$ CL. 

We have also studied the plausibility of the SMBH growth model against the observed SMBH population. We have found that the JWST population can be accommodated and that the best-fit parameters fall within the range of expected values from simulations, physical arguments and the PTA gravitational wave background strength. We also find that the feasibility of light seeds does not depend only on their ability to grow at super-Eddington rates (which is favoured but not required), but on whether they are planted in sufficiently light halos. These proto-galaxies will eventually be trapped in an overdensity, which will cause the small galaxies to merge into a bigger one. If the SMBHs that take part can merge efficiently enough, they can become viable seeding candidates. This makes the bounds on DM dependent on the seeding scenario, and light seeds provide more stringent bounds on deviations from CDM. This is an interesting phenomenon that could be explored with future gravitational wave detectors such as LISA~\cite{LISA:2024hlh} and AION~\cite{AION:2025igp}.

\vspace{5pt}\noindent\emph{Acknowledgments --} This work was supported by the Estonian Research Council grants PRG803, PSG869, RVTT3 and RVTT7 and the Center of Excellence program TK202. The work of J.E. and M.F. was supported by the United Kingdom STFC Grants ST/T000759/1 and ST/T00679X/1. The work of V.V. was partially supported by the European Union's Horizon Europe research and innovation program under the Marie Sk\l{}odowska-Curie grant agreement No. 101065736.

\bibliographystyle{JHEP.bst}
\bibliography{refs}

\newpage
\clearpage
\onecolumngrid

\begin{center}
\textbf{\large \papertitle} \\ 
\vspace{0.06in}
{John Ellis, Malcolm Fairbairn, Juan Urrutia and Ville Vaskonen} \\ 
\vspace{0.1in}
{SUPPLEMENTAL MATERIAL}
\vspace{0.1in}
\end{center}

\setcounter{equation}{0}
\setcounter{figure}{0}
\setcounter{section}{0}
\setcounter{table}{0}
\setcounter{page}{1}
\makeatletter
\renewcommand{\theequation}{S\arabic{equation}}
\renewcommand{\thefigure}{S\arabic{figure}}
\renewcommand{\thetable}{S\arabic{table}}

\section{Growth of galaxies and their central black holes}

To evolve the DM structures, galaxies and SMBHs, we follow the approach derived in~\cite{Ellis:2024nzv}, which is based on an empirical estimate of the star formation rate, a parametrization of the SMBH accretion rate that accounts for feedback effects, and the growth rate of galaxies and SMBHs estimated from the Extended Press-Schechter (EPS) formalism for an elliptical collapse barrier. The average BH mass and the stellar mass evolve according to
\bea
	&\dot M_{\rm BH}(M, p_{\rm BH}) = \dot M_{\rm BH}^{\rm merg.}(M, p_{\rm BH}) + \dot M_{\rm BH}^{\rm acc.}(M_{\rm BH},M) 
    \,, \\	
	&\dot M_*(M) = \dot M_*^{\rm merg.}(M) + \dot M_*^{\rm sf.}(M) \,.
\eea
The growth rate by mergers, $\dot M_J^{\rm merg.}(M)$, $J =\left( {\rm BH},\,*\right)$, in DM halos of mass $M$ is obtained from the limit $z'\to z$ of
\be \label{eq:mavevol}
    \Delta M_J^{\rm merg}(M,z,z') = \int_0^M \!\!\td M' M_J(M',z')\frac{\td N(M,z,z')}{\td M'} - M_J(M,z') \,,
\ee
where
\be
     \frac{\td N(M,z,z')}{\td M'} = \frac{M}{M'} \frac{\td S}{\td M'} p_{\rm FC}(M',z'|M,z) 
\ee
gives the number of progenitors of the halo of mass $M$ in the mass range $(M',M'+\td M')$. The function $S(M)$ is the variance of the matter fluctuations that depends on the DM model, and $p_{\rm FC}(M',z'|M,z)$ is the first-crossing probability distribution with $M > M'$ and $z' > z$ that we evaluate using the fit that we derived in~\cite{Ellis:2025xju}, which considers elliptical collapse and the DM dependence in the variance of matter perturbations. 

For the growth of the SMBHs via mergers, we also take into account that not all halo mergers lead to mergers of the SMBHs. For example, a third halo may merge with the system and bring a third SMBH before the SMBH pair merges. This may lead to ejection of one of the SMBHs. To account for this, we include a factor $p_{\rm BH}$ that characterises the probability that the BHs inside merging halos will themselves merge. On the other hand, if only one of the merging halos includes a SMBH, then the average SMBH mass in the resulting larger halos decreases. Consequently, the growth rate of average SMBH mass by mergers is given by
\be
    \dot M_{\rm BH}^{\rm merg}(M, p_{\rm BH}) = \!
    \begin{cases}
        \dot{M}_{\rm BH}^{\rm merg}(M), & \!\!\dot{M}_{\rm BH}^{\rm merg}(M)<0 \\
        p_{\rm BH} \,\dot{M}_{\rm BH}^{\rm merg}(M), & \!\!\dot{M}_{\rm BH}^{\rm merg}(M)\geq 0
    \end{cases} ,
\ee
where $\dot{M}_{\rm BH}^{\rm merg}(M)$ without argument $p_{\rm BH}$ is the growth rate obtained from Eq.~\eqref{eq:mavevol}. Although considering a constant $p_{\rm BH}$ might be simplistic, it can be used to approximate different merging mechanisms. For example, nuclear star clusters would lead to fast mergers and, consequently, $p_{\rm BH} \simeq 1$~\cite{Mukherjee:2024krx}.

For the star formation rate (SFR), we adopt the best fit that we found in~\cite{Ellis:2025xju}, derived from the HST and JWST UV luminosity function data. This fit is based on a parametrisation where the galaxy stellar mass grows proportional to the DM halo mass, $\dot M_* \propto \dot{M}$, with a proportionality factor that is a broken power-law as a function of the halo mass. The DM model dependence enters through the halo growth rate $\dot{M}$ that we evaluate using the expression that we derived in~\cite{Ellis:2025xju}. 

\begin{figure}
    \centering
    \includegraphics[width=0.5\columnwidth]{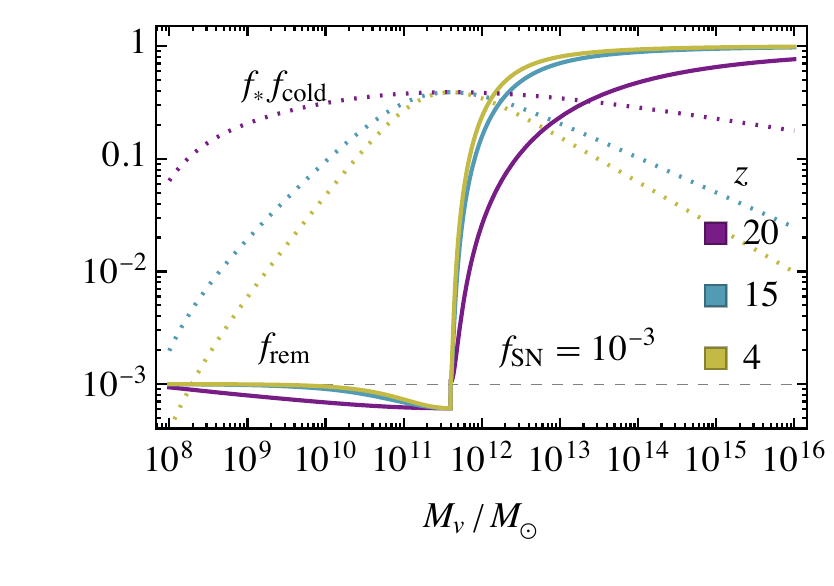}
    \vspace{-0.5 cm}
    \caption{The dotted curves in three colours corresponding to different redshifts show the SFR fitted to the UV luminosity function data, and the solid curves show the remaining gas fraction available for SMBH accretion at different redshifts.}
    \label{fig:feedbacks}
\end{figure} 

In order to estimate the SMBH accretion, it is essential to understand the feedback mechanisms within a galaxy. We infer these from the effective SFR by expressing it as $\dot M_* = f_*(M) f_{\rm cold}(M) f_B \dot{M}$ where $f_B = \Omega_B/\Omega_M \approx 0.16$ denotes the baryon fraction and $f_*(M) f_{\rm cold}(M)$ is a broken power-law that peaks at $M = M_c$, given by
\bea \label{eq:fstar}
    f_*\,f_{\rm cold}= \epsilon \,\frac{\alpha+\beta}{\beta(M/M_c)^{-\alpha}+\alpha(M/M_c)^{\beta}} \,e^{-M_t/M} .
\eea

We found in~\cite{Ellis:2025xju} that the best-fit values for these parameters do not depend significantly on the DM model, and here we use the best-fit values found in the CDM model. We divide the broken power-law into the fraction of cold gas 
\be
    f_{\rm cold}\equiv
    \begin{cases}
        1 \,, & M < M_c\\
        \dot M_*/(\epsilon f_B \dot M)\,, & M \geq M_c
    \end{cases} \,,
\ee
and the SFR efficiency
\be
    f_{*}\equiv
    \epsilon \begin{cases}
        \dot M_*/(\epsilon f_B \dot M) \,, & M < M_c\\
        1 \,, & M \geq M_c
    \end{cases} \,. 
\ee
The suppression of $f_*$ at $M<M_c$ arises because supernova (SN) feedback depletes the cold gas, halting further star formation, while the suppression of $f_{\rm cold}$ at $M>M_c$ is caused by AGN feedback, which heats the gas in the galaxy. According to this interpretation, the remaining gas that can be accreted by the SMBH is
\be
    f_{\rm rem} = 1 - (f_{\rm ej}+f_{*})f_{\rm cold} \,,
\ee
where 
\be
    f_{\rm ej} \equiv (1-f_{*}) (1-f_{\rm SN}) 
    \begin{cases}
        1 \,, & M < M_c\\
        f^2_{\rm cold} \,, & M \geq M_c
    \end{cases} \,.
\ee 
The quantity $f_{\rm SN}$ is the fraction of the gas left after star formation that is not expelled by the SN feedback. The motivation for introducing this parameter is to allow for some accretion on the SMBH at $M<M_c$, the halo mass which marks the transition from SN- to AGN-dominated feedback, and thus accommodate the low-mass AGNs seen by JWST. The solid curves in Fig.~\ref{fig:feedbacks} show the gas fraction available for SMBH accretion at different redshifts. Below $M_c$, SMBHs can only accrete the fraction of gas not expelled by SN feedback, which in the case shown in Fig.~\ref{fig:feedbacks} is $f_{\rm SN} = 10^{-3}$. Following~\cite{Dayal:2018gwg}, we consider the accretion rate
\be
    \dot M_{\rm BH}^{\rm acc}(M_{\rm BH},M) = \min \bigg[ f_{\rm Edd}\dot{M}_{\rm Edd}(M_{\rm BH}) f_{\rm rem} f_B (f_1^{\rm acc}\dot M + f_2^{\rm acc} M) \bigg] \,,
\ee
where $f_1^{\rm acc}$ is a dimensionless parameter that describes the fraction of baryons accreted by the halo that, after star formation and internal feedback, are ultimately accreted onto the SMBH. In contrast, $f_2^{\rm acc.}$ has units of inverse time and determines the rate at which the surrounding gas can be accreted. Finally, $f_{\rm Edd}$ is a dimensionless parameter that limits the total accretion rate as a fraction of the Eddington rate $\dot{M}_{\rm Edd}(M_{\rm BH})$. 

In the above formalism we evolve only the average BH mass, $M_{\rm BH} = \int \td P(m_{\rm BH}) \,m_{\rm BH}$. We assume that the occupation number is one as long as $M_{\rm BH} > m_{\rm seed}$ and in halos with $M_{\rm BH} < m_{\rm seed}$ the occupation number is less than one and the SMBH mass equals the seed mass. The occupation fraction can be expressed as
\be \label{eq:Pocc}
    \frac{\td P(m_{\rm BH}|M_*,z)}{\td m_{\rm BH}} = \frac{M_{\rm BH}(M_*,z)}{m_{\rm BH}} \delta\left[m_{\rm BH} - M_{\rm BH}(M_*,z)\right] \, . 
\ee

\newpage
\section{Posterior plots} 

\begin{figure*}[h!]
    \centering  \includegraphics[width=0.75\textwidth]{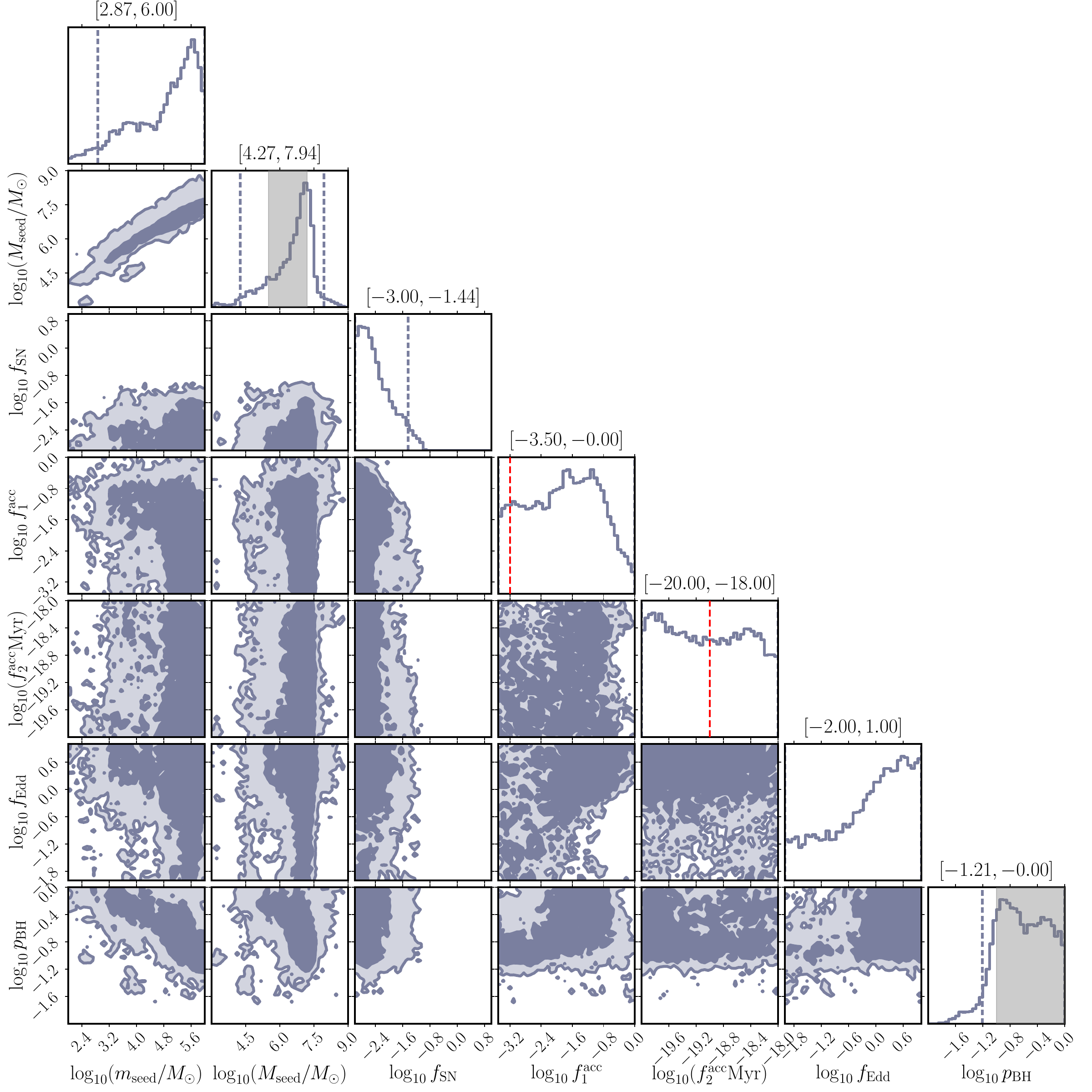}
    \caption{Posteriors of the stellar mass-BH mass relation fit in a CDM universe. The contours indicate the 68\% and 95\% credible regions. The grey dashed region in the $\log_{10}(M_{\rm seed}/M_{\odot})$ posterior represents the range of values given in the literature for this parameter~\cite{2007MNRAS.377..667N,2021MNRAS.507.1775S,2021ApJ...917...40K}, the two red dashed lines correspond to the values given in~\cite{2019MNRAS.486.2336D} to match the AGN luminosity function. The grey range on $p_{\rm BH}$ is the range that gives the correct PTA gravitational wave background strength according to~\cite{Ellis:2023dgf}.}
    \label{fig:scan_CDM}
\end{figure*}

\begin{figure*}[h!]
    \centering
\includegraphics[width=0.75\textwidth]{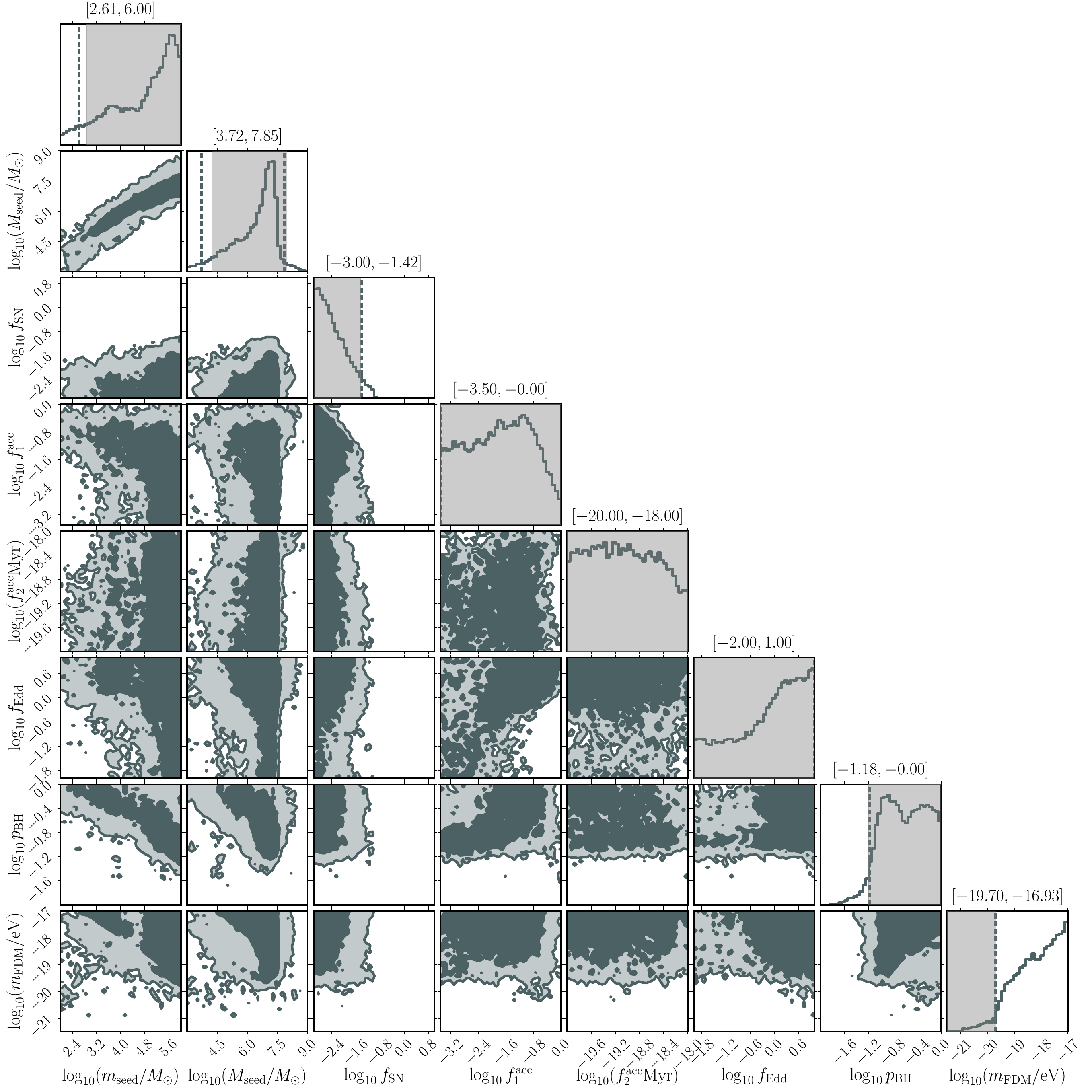}
    \caption{Same as Fig.~\ref{fig:scan_CDM} but for FDM universes. In the $\log_{10}(m_{\rm FDM}/{\rm eV})$ panel, the grey band marks the 95$\%\, {\rm CL}$ constraint on $m_{\rm FDM}$ from an analysis of Lyman-$\alpha$ data~\cite{Irsic:2017yje}, which excludes masses below $2\times10^{-20}\,{\rm eV}$. The grey bands show the $95\%$ CL ranges of the CDM fit.}
    \label{fig:scan_FDM}
\end{figure*}

\begin{figure*}[h!]
    \centering
\includegraphics[width=0.75\textwidth]{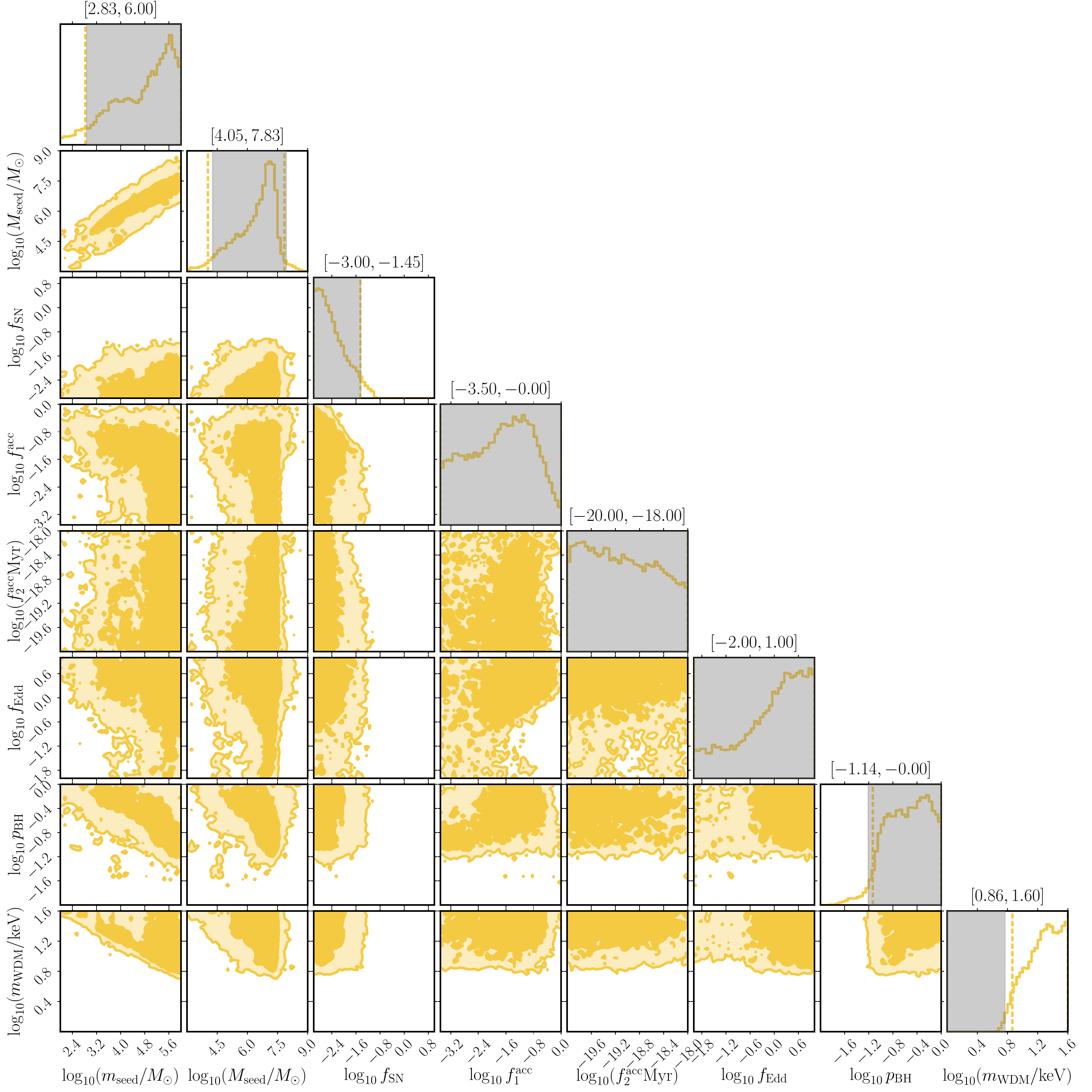}
    \caption{Same as Fig.~\ref{fig:scan_CDM} but for WDM universes. In the $\log_{10}(m_{\rm WDM}/{\rm keV})$ panel, the grey band marks the 95$\%\, {\rm CL}$ constraint on $m_{\rm WDM}$ from an joint analysis of strong gravitational lensing, the Lyman-$\alpha$ forest and Milky Way satellites~\cite{Enzi:2020ieg}, which excludes masses below $6.0\,{\rm keV}$. The grey bands show the $95\%$ CL ranges of the CDM fit.}
    \label{fig:scan_WDM}
\end{figure*}

\end{document}